\documentclass [12pt] {article}

\usepackage [utf8] {inputenc}
\usepackage [T1] {fontenc}
\usepackage {geometry}
\usepackage{amsmath}
\title {Excluded-volume effects of radial oscillations in disks confined to a circular box}
\author {James H. Taylor\\
School of Geoscience, Physics, and Safety\\
University of Central Missouri\\
Warrensburg, MO  64093\\
jtaylor@ucmo.edu}

\date {23 July 2019}

\begin {document}
\maketitle

\begin {abstract}
The effect of radial vibrations on the properties of one or two disks confined to a circular trap in contact with a  thermal reservoir are investigated. The vibrational amplitudes and energies are assumed to be quantized, with the motions corresponding roughly to certain modes for ringlike or tetrahedral molecules (such as benzene or methane, respectively). The calculation of the partition function requires integrations over the internal phases describing the oscillations, as well as the disks' center-of-mass positions and momenta; while an exact result is obtained for a single disk, for two disks the position-space integration can only be approximated. In spite of the small number of disks considered, various "thermodynamic" quantities are evaluated from the partition function. It is found that the average energy of the system is increased—compared to that for rigid disks—as are the entropy and compressibility; the pressure, however,  is decreased, and there is a variable effect on the heat capacity, depending on the ratio of the vibrational energy to the temperature. These changes can be traced either directly or indirectly to the oscillations causing an effective increase in the area available to the molecules within the circle, which in turn leads to an increase in the size of the accessible phase space, with the increase being larger for higher energy states. 
\\
\\
Rigorous results in statistical mechanics
\\
\\

\end {abstract}

\newpage
\begin{center}
I. INTRODUCTION
\end{center}

Since early work by Rayleigh \cite{ray} and Boltzmann \cite{bol} on corrections to the equation of state for a gas of finite molecules, numerous reseachers have considered the effect of size (and in some cases shape) on the properties of systems in one, two, or three dimensions. For example, Tonks \cite{ton} determined the equation of state for a "gas" of finite atoms on a planar substrate in both low- and high-density regimes, while---more recently---Grycko and Kirsch \cite{gry} obtained an approximate partition function for a hard-disk fluid in a rectangular box, using a theoretical analysis combined with input from a molecular dynamics simulation.

Olafsen and Urbach \cite{ola} investigated the horizontal velocity distribution and density fluctuations in a system of ball bearings on a vertically shaken horizontal plate, as did  Rouyer and Menon \cite{rou} for a system of bearings confined to a vertical plane. In both experiments, deviations from a Gaussian distribution were found under conditions where it was difficult or impossible for the "particles" to hop over each other, suggesting a possible excluded-volume origin. Working with bearings of two different sizes, shaken and confined in a vertical plane, Bose et al. \cite{bos} observed clustering of the larger ones, which they attributed to a pressure imbalance caused by the smaller species being prevented from moving between the larger when the latter are too close together. 

Monte Carlo studies of "conducting sticks" in 2D (Balberg and Binenbaum \cite{bbo}) and cylinders in 3D (Balberg et al. \cite{bbn}) elucidated the effects of size on percolation, while Zhang et al. \cite{zha} have shown how exclusion affects orientational ordering of a monolayer of hard rods on a spherical surface. The influence of exclusion has also been analyzed for long-chain molecules in 2D---both experimentally (Maier and R\"adler \cite{mai}, Drube et al. \cite{dru}) and via compuer simulations (Meyer et al. \cite{mey}, Hsu et al. \cite{hsu}, Drube et al \cite{dru}.)---, a double layer of ions near a charged wall (Frydel and Levin \cite{fry}), shear thickening (Madraki et al. \cite{mad}), and diffusion (Bruna and Chapman \cite{bru}). Onsager \cite{ons} considered the effect of shape on the interaction between colloidal particles; more recently, the particular impact of molecular geometry and orientation have been investigated in the contexts of chemical potential (Krukowski et al. \cite{kru}) and the transfer of vibrational energy between molecules in liquid water (Yang \cite{yan}).

The system considered here is composed of circular disks ("molecules") with an oscillating radius, confined to a circular space. (The corresponding case in 1D---a sytem of oscillating rods---has been examined previously. \cite{tay}) The oscillations approximate, for example, the (radially symmetric) A1-type vibrations of benzene or, more roughly, of BF$_{3}$, and may even be considered a crude description for methane in the A1 mode, with three of the hydrogens resting on a plane. In general agreement with earlier work, it is found that the effects are unimportant for a circle much larger than the disks, but significant when its radius is sufficiently small. 

The model is described in greater detail in the section immediately below. Sec. III involves the evaluation of the partition function for both a single disk and a pair of disks, and in Sec. IV this function is used to evaluate several "thermodynamic" properties. Brief conclusions follow in Sec. V. 
\\
\\
\begin{center}
II. THE MODEL
\end{center}

The disks each have a mass $M$, are assumed to be indistinguishable, and interact with each other---or with the boundary of radius $R$---only through collisions.  They undergo radial vibrations at a single angular frequency $\omega$, with the allowed energies being those of a quantum-mechanical harmonic oscillator. In the absence of oscillations, they would all have the same radius, $r_0$; if disk $n$ is in state $j$, the amplitude of the vibrations around $r_0$ will be designated by $\Delta_{r(j;n)}$, with $j=0$ corresponding to the ground state.  $\Delta_{r(j;n)}=\Delta_{r0}\sqrt{2j+1}$, where the ground-state amplitude $\Delta_{r0}=\sqrt{\hbar\omega/k_r}$, $k_r$ being an effective "spring constant." In spite of the assumption that the vibrational energy levels are discrete, the oscillations themselves will be treated classically. Finally, for the sake of simplicity, only states with $j=0$ or $1$ will be included; this is not unrealistic, since for the molecules mentioned in Sec. I the excitation energies correspond to a temperature of well over 1000 K.

The radius of disk $n$ is given by
\begin{equation}
r_n=r_0+\Delta_{r(j;n)}sin(\omega t+\theta_n)=r_0+\Delta_{r(j;n)}sin\delta_n,
\end{equation}
with $\theta_n$ being a random phase at some arbitrary time zero, corresponding to a random "starting time" for a particular vibrational state of that disk. It is assumed that the $\theta_n$'s for different disks are mutually independent, which means the $\delta_n$'s will be independent as well. In what follows, it will be more convenient to work with the latter.

Since they cannot spatially overlap, the allowed range of positions for a given disk depends on the positions and radii of the other disks, as well as on its own radius. It is this restriction that gives rise to effects directly dependent on both the "undisturbed" radius $r_0$ and the amplitude of the vibrations. 
\\
\\
\begin{center}
III. THE PARTITION FUNCTION
\end{center}

In the following, $Z_{dN}$ will be used to indicate the partition function for $N$ disks. (The subscript $d$ serves to help avoid confusion with corresponding results for the previously mentioned system of vibrating rods in 1D, which will be referenced later.) This may be expressed as a product of two terms: $Z^\textrm{p}_{dN}$, which depends only on the disks' momenta (the "momentum part"), and $Z^\textrm{x}_{dN}$ (the "position part"), which includes the vibrational part of the disks' energies, since this will have an effect on each disk's size, and thus on the positions available to them within the the circle.

Assuming the system is in contact with a thermal reservoir at temperature $T$, the momentum part of the partition function is given by
\begin{equation}
\frac{1}{h^{2N}}\prod_{n=1}^N\int^{\infty}_0\int^{2\pi}_0\textrm{e}^{-p_n^2/2Mk_\textrm{B}T}p_n\textrm{d}p_n\textrm{d}\phi_{pn}=\bigg(\frac{2\pi Mk_\textrm{B}T}{h^2}\bigg)^N,
\end{equation}
where $p_n$ is the magnitude of the center-of-mass momentum of disk $n$, $\phi_{pn}$ is the direction angle of that momentum in polar coordinates, $k_B$ is Boltzmann's constant, and $h$ is Planck's constant. 

While obtaining $Z^\textrm{p}_{dN}$ for an arbitrary numbers of disks is simple, the same is not true for $Z^\textrm{x}_{dN}$ when $N>1$. Only for $N=1$ can the position part be found exactly; the derivation of the result is presented in Part A below. Part B gives an approximate evaluation for $N=2$.
\\
\\
A. Evaluation of  $Z^\textrm{x}_{d1}$ 

The center of the disk can be no closer to the edge of the circle than its own instantaneous radius, $r_n$. Placing the coordinate origin at the center of the circle, and assuming its radius is greater than the maximum radius of the disk in the excited state, the position part of the partition function is 
\begin{equation}
Z^\textrm{x}_{d1}=\sum_{(j;1)=0}^1\textrm{e}^{-E_1^{\textrm{int}}/k_\textrm{B}T}\int^{2\pi}_0\textrm{d}\delta_1\bigg\lbrace\int^{R-r_0-\Delta_{r(j;1)}\textrm{sin}\delta_1}_0\int^{2\pi}_0\rho_1\textrm{d}\rho_1\textrm{d}\phi_1   \bigg\rbrace,
\end{equation}
where $E_1^{\textrm{int}}$ is the internal vibrational energy. Integrating first over the position coordinates and then over the phase $\delta_1$ yields
\begin{equation}
Z^\textrm{x}_{d1}=\sum_{(j;1)=0}^1\textrm{e}^{-E_1^{\textrm{int}}/k_\textrm{B}T}\bigg[(R-r_0)^2+\frac{\Delta_{r(j;1)}^2}{2}\bigg].
\end{equation}
Summing over $(j;1)$ and multiplying by $Z^\textrm{p}_{d1}$, one finally obtains
\begin{multline}
Z_{d1}=\bigg(\frac{\pi Mk_\textrm{B}T}{\hbar^2}\bigg)q^{1/2}(1+q)R^2\bigg[(1-r_0^\prime)^2+\frac{r_0^{\prime2} (\Delta_{r0}^\prime)^2}{2}\bigg(\frac{1+3q}{1+q}\bigg)\bigg]
\\=\bigg(\frac{\pi Mk_\textrm{B}T}{\hbar^2}\bigg)q^{1/2}(1+q)R^2Y_{d1}
\end{multline}
where $r_0^\prime=(r_0/R)$ and $\Delta_{r0}^\prime=(\Delta_{r0}/r_0)$ (both presumably small under normal circumstances), and the quantities $q=\textrm{e}^{-\hbar\omega/k_\textrm{B}T}$ and $Y_{d1}$ have been introduced for the sake of notational brevity. Note that increasing the value of $r_0$ causes a decrease in $Y_{d1}$ (a direct result of the decrease in available area in position space), whereas an increase in $\Delta_{r0}$ has the opposite effect.

In the aforementioned system of oscillating rods, the partition function for a single "molecule" had no dependence on the vibrational amplitude. In the present case, the appearance of $\Delta_{r0}^\prime$ can perhaps be understood qualitatively (using Cartesian spatial coordinates) by recognizing that the range of allowed positions for the disk's center along the y axis is restricted by the position of the center along the x axis (and vice versa). In this sense, the situation is similar to that of two rods oscillating in one dimension, with the instantaneous length of each rod affecting the range of possible locations for the other. In fact, the corresponding result is remarkably similar:
\begin{equation}
Z_{21}=\bigg(\frac{\pi Mk_\textrm{B}T}{\hbar^2}\bigg) q(1+q)^2L^2\bigg[(1-2l^{\prime})^2+4l^{\prime2}(\Delta_{02}^{\prime})^2\bigg(\frac{1+3q}{1+q}\bigg)\bigg],
\end{equation} 
with $l^\prime=(l/L)$ and $\Delta^\prime_{02}=(\Delta_{0}/l)$ ($l$ is the length of a rod with no vibrations, $2\Delta_{0}$ is the ground-state amplitude of the change in length, and $L$ is the length of the system; the subscript $1$ indicates that only the ground and first excited states are included in the sum). The principal difference between the two expressions is the factor of $q^{1/2}(1+q)$ in equation (5) vs. the square of the same in equation (6), which is of course due to the different number of oscillators. 
\\
\\B. Approximate evaluation of $Z^\textrm{x}_{d2}$

We can begin by writing the position part of  $Z_{d2}$ in the form   
\begin{equation}
Z^\textrm{x}_{d2}=\frac{1}{2!}\sum_{(j;1)=0}^1\sum_{(j;2)=0}^1\textrm{e}^{-(E_1^{\textrm{int}}+E_2^{\textrm{int}})/k_\textrm{B}T}\int^{2\pi}_0\int^{2\pi}_0\textrm{d}\delta_1 \textrm{d}\delta_2(I_1+I_2), 
\end{equation}
where the factorial accounts for the indistinguishability of the disks, and the quantity $(I_1+I_2)$ represents the integral over the possible combinations of positions of the two centers. To evaluate this integral, first note that: $(1)$ neither center can be closer to the boundary than that disk's instantaneous radius; and $(2)$ that the distance between centers cannot be smaller than the sum of the radii, $(r_1+r_2)$. In the following, it will be arbitrarily assumed that disk 1 is free to range anywhere within the circle that is allowed by the first requirement, with the possible locations for disk 2 then restricted by the actual location of disk 1.

Two different situations can be considered separately: either the center of disk 1 is closer to the boundary than $(r_1+2r_2)$, or it is not. In the latter case (which will correspond to $I_1$), disk 2 is free to make a complete circle around disk 1 without touching the boundary, and it is unnecessary to go through the formal process of integration: the magnitude of the area accessible to its center is simply $\pi(R-r_2)^2$ (the area available to it if disk 1 were not present) minus $\pi(r_1+r_2)^2$ (the excluded area around the center of disk 1). To complete this part of the combined position integration, it is only necessary to multiply by $\pi(R-r_1-2r_2)^2$, the area open to disk 1 given the stipulation on its distance from the boundary. Thus,
\begin{equation}
I_1=\pi^2(R-r_1-2r_2)^2[(R-r_2)^2-(r_1+r_2)^2].
\end{equation}

The more complicated case occurs when disk 1 is too close to the boundary for disk 2 to fit between them. In this situation (which will correspond to $I_2$), the area from which disk 2 is excluded is actually smaller than before: now, part of the prohibited circle around disk 1 lies outside the circle of radius $(R-r_2)$.

First, the overlap between the two circles just mentioned is subtracted from $\pi(R-r_2)^2$, then the result is integrated over all possible positions of the center of disk 1 with radial coordinate $\rho_1$ between $(R-r_1-2r_2)$ and $(R-r_1)$: 
\begin{multline}
\int^{2\pi}_0\int^{(R-r_1)}_{(R-r_1-2r_2)}\rho_1\textrm{d}\rho_1\textrm{d}\phi_1 \bigg\lbrace\frac{1}{2}\sqrt{4\rho_1^2(R-r_2)^2-[\rho_1^2+(R-r_2)^2-(r_1+r_2)^2]^2}
\\+\pi(R-r_2)^2-(r_1+r_2)^2\textrm{cos}^{-1}\bigg[\frac{\rho_1^2+(r_1+r_2)^2-(R-r_2)^2}{2\rho_1 (r_1+r_2)}\bigg]
\\-(R-r_2)^2\textrm{cos}^{-1}\bigg[\frac{\rho_1^2+(R-r_2)^2-(r_1+r_2)^2}{2\rho_1(R-r_2)}\bigg]\bigg\rbrace.      
 \end{multline} 

 The integration over $\phi_1$ is trivial. Following that, evaluation of the first two terms is straightforward. The third term can be written as
\begin{equation}
-(r_1+r_2)^2\int^{(R-r_1)}_{(R-r_1-2r_2)}vdu
\end{equation}
where $\textrm{d}u=\rho_1 \textrm{d}\rho_1$ and $v$ is the rest of the integrand; the fourth term can be expressed similarly. After integrating by parts, defining the argument of the inverse cosine as a new variable reduces the remaining integrals to a more manageable form. Combining all the pieces and making use of several identities involving inverse trigonometric functions \cite{gra}, we obtain
\begin{multline}
I_1+I_2=2\pi[R^2-R(r_1+r_2)+(r_1^2+r_2^2+r_1r_2)]\sqrt{r_1r_2R(R-r_1-r_2)}
\\+2\pi(R-r_2)(R^2-r_2^2)(R-2r_1-r_2)\textrm{cos}^{-1}\bigg[\sqrt{\frac{r_1R}{(R-r_2)(r_1+r_2)}}\bigg]
\\+2\pi(R-r_1)(R^2-r_1^2)(R-r_1-2r_2)\textrm{cos}^{-1}\bigg[\sqrt{\frac{r_2R}{(R-r_1)(r_1+r_2)}}\bigg].
\end{multline}
Note that this result is symmetric under the interchange of $r_1$ and $r_2$, as it must be.

The next step is to integrate over all possible values of the phases $\delta_1$ and $\delta_2$. Because $r_n=r_0+\Delta_{r(j;n)}\textrm{sin}\delta_n$, it appears that the integrals cannot be evaluated exactly, so we approximate both the inverse cosines and the square root as power series. As for a single disk, the final results will be expressed in powers of the two small quantities, $r_0^\prime$ and $\Delta_{r0}^\prime$.

In order to expand the inverse cosines, it will be more convenient to eliminate the square roots by noting that $2\textrm{cos}^{-1}x = \textrm{cos}^{-1}(2x^2-1)$ \cite{gra}, leading to
\begin{multline}
I_1+I_2=2\pi[R^2-R(r_1+r_2)+(r_1^2+r_2^2+r_1r_2)]\sqrt{r_1r_2R(R-r_1-r_2)}
\\+\pi(R-r_2)(R^2-r_2^2)(R-2r_1-r_2)\textrm{cos}^{-1}\bigg[\frac{2r_1R}{(R-r_2)(r_1+r_2)}-1\bigg]
\\+\pi(R-r_1)(R^2-r_1^2)(R-r_1-2r_2)\textrm{cos}^{-1}\bigg[\frac{2r_2R}{(R-r_1)(r_1+r_2)}-1\bigg].
\end{multline}
Using \cite{gr2} 
\begin{equation}
\textrm{cos}^{-1}x=\frac{\pi}{2}-\sum_{k=0}^\infty\frac{(2k)!}{2^{2k}(k!)^2(2k+1)}x^{2k+1},
\end{equation}
as well as standard series approximations for $(1+y)^{1/2}$ and $(1+y)^{-1}$ when $y\ll1$, and retaining terms up to sixth order of smallness, we obtain, after a good deal of tedious algebra,
\begin{multline}
I_1+I_2\approx\pi^2R^4\lbrace1-4r_0^\prime+2r_0^\prime[r_0^\prime-2B_+]+4r_0^{\prime2}[r_0^\prime+B_+]
\\+r_0^{\prime2}[-3r_0^{\prime2}+12r_0^\prime B_++2(B_+^2-B_-^2)]+4r_0^{\prime3}[-3r_0^\prime B_++(3B_+^2+B_-^2)]
\\+2r_0^{\prime3}[-3r_0^\prime(3B_+^2+B_-^2)+2(B_+^3+B_-^2B_+)]\rbrace
\\+\pi R^4\lbrace\frac{32}{3}r_0^{\prime3}+32r_0^{\prime3}[-\frac{1}{3}r_0^\prime+B_+]+16r_0^{\prime3}[-\frac{1}{15}r_0^{\prime2}-\frac{8}{3}r_0^\prime B_++(2B_+^2+B_-^2)]
\\+16r_0^{\prime3}[-\frac{1}{15}r_0^{\prime3}-\frac{1}{3}r_0^{\prime2}B_+-r_0^\prime(4B_+^2+B_-^2)+(\frac{2}{3}B_+^3+B_-^2B_+)]\rbrace,
\end{multline}
where 
\begin{equation}
B_{\pm}=\frac{\Delta_{r2}\textrm{sin}\delta_2\pm \Delta_{r1}\textrm{sin}\delta_1}{2r_0}.
\end{equation}

Finally integrating over the phases gives
\begin{multline}
I_1+I_2\approx4\pi^4R^4\bigg\lbrace1-4r_0^\prime+2r_0^{\prime2}+4r_0^{\prime3}\bigg[1+\frac{(\Delta_{r1}^2+\Delta_{r2}^2)}{2r_0^2}\bigg]-3r_0^{\prime4}\bigg[1+\frac{(\Delta_{r1}^2+\Delta_{r2}^2)}{r_0^2}\bigg]\bigg\rbrace
\\+8\pi^3R^4\bigg\lbrace r_0^{\prime3}\bigg[\frac{16}{3}+\frac{3(\Delta_{r1}^2+\Delta_{r2}^2)}{r_0^2}\bigg]-r_0^{\prime4}\bigg[\frac{16}{3}+\frac{5(\Delta_{r1}^2+\Delta_{r2}^2)}{r_0^2}\bigg]-\frac{8}{15}r_0^{\prime5}-\frac{8}{15}r_0^{\prime6}\bigg\rbrace.
\end{multline}

Inserting this into equation (7), carrying out the summations, and multiplying by $Z^\textrm{p}_{d2}$  yields
\begin{multline}
Z_{d2}\approx\frac{1}{2}\bigg(\frac{\pi Mk_\textrm{B}T}{\hbar^2}\bigg)^2q(1+q)^2R^4\bigg(\bigg\lbrace(1-3r_0^\prime)(1+r_0^\prime)(1-r_0^\prime)^2
\\+\frac{16}{3\pi}r_0^{\prime3}\bigg[2(1-r_0^\prime)-\frac{1}{5}r_0^{\prime2}(1+r_0^{\prime})\bigg]\bigg\rbrace+2r_0^{\prime3}(\Delta'_{r0})^2\bigg(\frac{1+3q}{1+q}\bigg)\bigg\lbrace(2-3r_0^\prime)+\frac{2}{\pi}(3-5r_0^\prime)\bigg\rbrace\bigg).
\end{multline}

While the effect of the oscillations is first made manifest at fourth order of smallness in $Z_{d1}$, the first term in this expression showing dependence on $\Delta_{r0}$ is of fifth order: each of the three parts in equation (12), when approximated and integrated over the phases, includes terms proportional to both $r_0^{\prime}(\Delta'_{r0})^2$ and $(r_0^{\prime}\Delta'_{r0})^2$, but they exactly cancel, and there are no terms proportional to $(\Delta'_{r0})^2$. It is unclear whether or not such lower-order terms would appear for a system containing more than two disks.  
\\
\\
\begin{center}
IV. SYSTEM PROPERTIES
\end{center}

\begin{center}
A. Energy and constant-area heat capacity 
\end{center}

In the present notation, the average energy of a system containing N molecules at temperature $T$ is
\begin{equation}
E_{N} = \frac{k_\textrm{B}T^2}{Z_{dN}}\bigg( \frac{\partial Z_{dN} }{\partial T}\bigg), 
\end{equation} 
and the corresponding heat capacity at constant area is given by
\begin{equation}
C_{AN} = \bigg( \frac{\partial E_{N}}{\partial T} \bigg)_A.
\end{equation}
For $N = 1$, equation (18) gives
\begin{equation}
E_1=k_\textrm{B}T+\frac{\hbar\omega}{2}+\hbar\omega\bigg(\frac{q}{1+q}\bigg)+\hbar\omega\bigg(\frac{q}{1+q}\bigg)\bigg[\frac{ r_0^{\prime2}(\Delta'_{r0})^2}{(1+q)Y_{d1}}\bigg].
\end{equation}
The origin of the first two terms is obvious; the third is the increase in the average vibrational energy due to $T$ being greater than zero. Only the fourth term is affected by either the amplitude of the vibrations or the molecular size. Clearly, $E_1$ is larger when $\Delta_{r0}$ is greater than zero, and closer inspection shows that it increases monotonically with the amplitude. A quick look at equation (4) reveals that the increase in the magnitude of the oscillations with increasing energy means a molecule in the excited state has a larger average area in position space available to it than it would in the ground state. The resulting increase in the available phase-space volume leads to a greater relative probability for the state with higher energy than it would have without oscillations, and thus an increase in the average total energy. 

The heat capacity is
\begin{equation}
C_{A1}=k_\textrm{B}+k_\textrm{B}\bigg(\frac{\hbar\omega}{k_\textrm{B}T}\bigg)^2\frac{q}{(1+q)^2}\bigg[1+\frac{r_0^{\prime2}(\Delta'_{r0})^2}{(1+q)Y_{d1}}\bigg]\bigg[1-\frac{qr_0^{\prime2}(\Delta'_{r0})^2}{(1+q)Y_{d1}}\bigg].
\end{equation}
This is increased by $\Delta_{r0}>0$ only if
\begin{equation}
q^2<\frac{2(1-r_0^\prime)^2+r_0^{\prime2}(\Delta_{r0}^\prime)^2}{2(1-r_0^{\prime})^2+3r_0^{\prime2}(\Delta_{r0}^\prime)^2}.
\end{equation}
Given the definition of q, this implies that $C_{A1}$ is raised by the oscillations at low $T$, but lowered by them when the temperature is sufficiently high.

The expressions for the energy and heat capacity have forms very similar to those for two rods oscillating in 1D \cite{tay}. 

When $N=2$, one obtains, to sixth order in small terms,
\begin{multline}
E_2\approx2k_\textrm{B}T+\hbar\omega+2\hbar\omega\bigg(\frac{ q}{1+q}\bigg)+
\\4\hbar\omega\bigg(\frac{q}{1+ q}\bigg)\bigg[\frac{r_0^{\prime3}(\Delta_{r0}^\prime)^2}{1+q}\bigg]\bigg[2\bigg(1+\frac{3}{\pi}\bigg)+\bigg(5+\frac{14}{\pi}\bigg)r_0^\prime\bigg]
\end{multline}
and
\begin{multline}
C_{A2}\approx2k_\textrm{B}+2k_\textrm{B}\bigg(\frac{\hbar\omega}{k_\textrm{B}T}\bigg)^2\frac{q}{(1+q)^2}\bigg\lbrace1-\frac{2(1-q)}{(1+q)} r_0^{\prime3}(\Delta_{r0}^\prime)^2\bigg[2\bigg(1+\frac{3}{\pi}\bigg)
\\+\bigg(5+\frac{14}{\pi}\bigg)r_0^\prime\bigg]\bigg\rbrace.
\end{multline}
Since $q$ is always less than 1, $C_{A2}$ is \textit{reduced} if $\Delta_{r0}>0$.

\begin{center}
B. Entropy
\end{center}

The entropy $S_{N} = E_{N}/T+k_\textrm{B}\textrm{ln}(Z_{dN})$, which for $N=1$ yields
\begin{multline}
S_1=k_\textrm{B}+k_\textrm{B}\textrm{ln}\bigg[\bigg(\frac{\pi Mk_\textrm{B}T}{\hbar^2}\bigg)(1+q)R^2Y_{d1}\bigg]
\\+k_\textrm{B}\bigg(\frac{\hbar\omega}{k_\textrm{B}T}\bigg) \bigg(\frac{q}{1+q} \bigg)\bigg\lbrace1+\bigg[\frac{r_0^{\prime2}(\Delta'_{r0})^2}{(1+q)Y_{d1}}\bigg]\bigg\rbrace,
\end{multline}
whereas for two disks, we find
\begin{multline}
S_2\approx2k_\textrm{B}+2k_\textrm{B}\textrm{ln}\bigg[\frac{1}{2}\bigg(\frac{\pi Mk_\textrm{B}T}{\hbar^2}\bigg)(1+q)R^2\bigg]-k_\textrm{B}r_0^\prime\bigg[4+6r_0^\prime+\bigg(\frac{28}{3}-\frac{32}{3\pi}\bigg)r_0^{\prime2}
\\+\bigg(21-\frac{32}{\pi}\bigg)r_0^{\prime3}+\bigg(\frac{244}{5}-\frac{528}{5\pi}\bigg)r_0^{\prime4}+\bigg(122-\frac{944}{3\pi}+\frac{512}{9\pi^2}\bigg)r_0^{\prime5}\bigg]
\\+2k_\textrm{B}\bigg(\frac{1+3q}{1+q}\bigg)r_0^{\prime3}(\Delta_{r0}^\prime)^2\bigg[2\bigg(1+\frac{3}{\pi}\bigg)+\bigg(5+\frac{14}{\pi}\bigg)r_0^\prime\bigg]
\\+2k_\textrm{B}\bigg(\frac{\hbar\omega}{k_\textrm{B}T}\bigg)\bigg(\frac{q}{1+q}\bigg)\bigg\lbrace1+\bigg(\frac{2}{1+q}\bigg)r_0^{\prime3}(\Delta_{r0}^\prime)^2\bigg[2\bigg(1+\frac{3}{\pi}\bigg)+\bigg(5+\frac{14}{\pi}\bigg)r_0^\prime\bigg]\bigg\rbrace.
\end{multline}
In both cases, if $\Delta_{r0}$ is held fixed, finite disk size \textit{reduces} the entropy, whereas the effect of finite oscillations at constant $r_0^\prime$ is to \textit{increase} it. This to be expected because of the way each factor affects the size of the available area in position space.

\begin{center}
C. Two-dimensional pressure and isothermal compressibility
\end{center}

For a general system of $N$ molecules with a single excited state, the pressure in this circular space would be
\begin{equation}
P^{2D}_N=  \frac{k_\textrm{B}T}{Z_{dN}}\bigg(\frac{\partial Z_{dN}}{\partial A}\bigg)=\frac{k_\textrm{B}T}{2\pi RZ_{dN}}\bigg(\frac{\partial Z_{dN}}{\partial R}\bigg),
\end{equation}
where $\pi R^2$ has been used for the area $A$.

For a single molecule, we obtain
\begin{equation}
P^{2D}_1=\bigg(\frac{k_\textrm{B}T}{\pi R^2}\bigg)\bigg[ \frac{(1-r_0^{\prime})}{Y_{d1}}\bigg],
\end{equation}
while for two molecules we get,
\begin{multline}
P^{2D}_2\approx\bigg(\frac{2k_{\textrm{B}}T}{\pi R^2}\bigg)\bigg\lbrace1+r_0^\prime+3r_0^{\prime2}+\bigg[7-\frac{8}{\pi}\bigg]r_0^{\prime3}+\bigg[21-\frac{32}{\pi}\bigg]r_0^{\prime4}+\bigg[61-\frac{132}{\pi}\bigg]r_0^{\prime5}
\\+\bigg[183-\frac{472}{\pi}+\frac{256}{3\pi^2}\bigg]r_0^{\prime6}-\bigg(\frac{1+3q}{1+q}\bigg)r_0^{\prime3}(\Delta_{r0}^\prime)^2\bigg[3\bigg(1+\frac{3}{\pi}\bigg)+2\bigg(5+\frac{14}{\pi}\bigg)r_0^\prime\bigg]\bigg\rbrace.
\end{multline}
As with $S_N$, we see $r_0\neq0$ and $\Delta_{r0}\neq0$ producing opposing effects: increasing the (unperturbed) size of the molecules raises the pressure, whereas increasing the amplitude of the oscillations lowers it.

Finally, the isothermal compressibility is obtained from
\begin{equation}
\kappa^{2D}_{T,N}=-\frac{1}{A}\bigg(\frac{\partial A}{\partial P^{2D}_{N}}\bigg)_T.
\end{equation}
For a single disk, this gives us
\begin{equation}
\kappa^{2D}_{T,1}=\bigg(\frac{\pi R^2}{k_\textrm{B}T}\bigg)\bigg[\frac{Y_{d1}^2}{(1-r_0^{\prime})^2-\frac{1}{2}r_0^{\prime}Y_{d1}}\bigg].
\end{equation}
Closer inspection indicates, unsurprisingly, that $\kappa^{2D}_{T,1}$ is greater when $\Delta_{r0}$ is nonzero.

For two disks we find, to sixth order,
\begin{multline}
\kappa^{2D}_{T,2}\approx\bigg(\frac{\pi R^2}{2k_{\textrm{B}T}}\bigg)\bigg\lbrace1-\frac{3}{2}r_0^\prime-\frac{15}{4}r_0^{\prime2}-\bigg[\frac{23}{8}-\frac{20}{\pi}\bigg]r_0^{\prime3}-\bigg[\frac{159}{16}-\frac{36}{\pi}\bigg]r_0^{\prime4}
\\ -\bigg[\frac{679}{32}-\frac{69}{\pi}\bigg]r_0^{\prime5}-\bigg[\frac{2159}{64}-\frac{108}{\pi}-\frac{176}{3\pi^2}\bigg]r_0^{\prime6}
\\+\bigg(\frac{1+3q}{1+q}\bigg)r_0^{\prime3}(\Delta_{r0}^\prime)^2\bigg[\frac{15}{2}\bigg(1+\frac{3}{\pi}\bigg)+\frac{3}{2}\bigg(5+\frac{11}{\pi}\bigg)r_0^\prime\bigg]\bigg\rbrace.
\end{multline}
\\
\\
As for the single disk, the compressibility is greater than it would be if $\Delta_{r0}$ were zero, and once more we see that the molecular size and finite oscillation-amplitude have opposite effects. 
\\
\\
\begin{center}
V. CONCLUSIONS
\end{center}

For a 2D system composed of one or two finite, disk-like "molecules" with sinsuoidally oscillating radius, changes in a variety of thermodynamic quantities are found, compared to the point-particle case. Corrections to the energy, entropy, pressure, etc. depend on the ratio of the vibrational energy to the temperature, and are found to be relatively small when $\Delta_{r0}\ll (R-r_0)$ (which would normally be the case). As could be expected, in the absence of vibrations the finite size of the disks had no effect on the energy or constant-area heat capacity of the system, but lowered the entropy and compressibility, while raising the pressure. On the other hand, vibrations did affect the energy and heat capacity, and had the opposite effect of a fixed, finite radius on the other properties just mentioned. Only even powers of the vibartional amplitude appear in any of the quantities considered, which is simply attributable to the fact that averaging an odd power of a sinsuodial function over a complete cycle yields zero.   

While it does not seem feasible to extend the analysis undertaken here to a system of three or more disks, a study of the behavior of larger systems may be possible via a molecular dynamics approach. 
\\

\end {document}